\documentclass{article}%
\usepackage{amsfonts}
\usepackage{amsmath}
\usepackage[doublespacing]{setspace}
\usepackage{graphicx}
\usepackage{amssymb}%
\providecommand{\U}[1]{\protect\rule{.1in}{.1in}}

\begin{document}

\begin{center}
Relationship between ferroelectricity and Dzyaloshinskii-Moriya interaction in
multiferroics and the effect of bond-bending

C. D. Hu

Department of Physics and Center for Theoretical Sciences

National Taiwan University

Taipei, Taiwan

Republic of China

\bigskip

Abstract

\medskip
\end{center}

We studied the microscopic mechanism of multiferroics, in particular with the
"spin current" model (H. Katsura, N. Nagaosa and A. V. Balatsky, Phys. Rev.
Lett. \textbf{95}, 057205 (2005)). Starting from a system with helical spin
configuration, we solved for the forms of the electron wave functions and
analyzed their characteristics. The relation between ferroelectricity and
Dzyaloshinskii-Moriya interaction (I. Dzyaloshinskii, J. Phys. Chem. Solids
\textbf{4}, 241 (1958) and T. Moriya, Phys. Rev. \textbf{120}, 91 (1960)) is
clearly established. There is also a simple relation between the electric
polarization and the wave vector of magnetic orders. Finally, we show that the
bond-bending existing in transition metal oxides can enhance ferroelectricity.

\noindent PACS: 75.80.+q, 77.80.-e\newpage

\subsubsection{1. Introduction}

Experimental findings[1-4] in recent years have revived the interest in
multiferroics. They showed that magnetic and ferroelectric orders are closely
related[5-8]. What is more intriguing is that only certain types of magnetic
orders, namely helical spins and frustrated spins, can be coupled to
ferroelectricity[9]. It is this fascinating interplay between ferroelectric
and magnetic orders that has attracted many researchers. There were already
models based on the Ginzburg-Landau theory[10,11] that provide instructive
physical description of the systems. As for the microscopic mechanism, there
are currently two schools of theories. One of them proposed that the electric
polarization and the anomaly of dielectric constant come from atomic
displacements. The displacements or phonons are in turn, coupled to
spins[12-14]. Though proposed for systems of orthorhombic structure, it is
more readily applied to multiferroics of hexagonal structures, such as
HoMnO$_{3}$, as there is experimental evidence of atomic displacements from
neutron scattering data[15]. The second school of theory proposed a new
possibility: electric polarization coming from electronic wave function and
thus density distribution. Katsura et. al.(KNB)[16] predicted that the
magnetoelectric effect can be induced by "spin current"[17]. The coupling
between "spin current" and internal electric field has the same form as that
of Dzyaloshinskii-Moriya interaction (DM)[18,19] or AC-effect[20] where the
motion of a magnetic moment is coupled to electric field. In this latter
theory, the atomic displacement is not essential. On the other hand,
spin-orbit interaction is indispensable in generating electric dipole moments.

The "spin current" model, though a bright idea, needs additional
substantiation in order to be applied to physical systems. Jia[21] et. al.
gave a detailed calculation of this model. Their results showed that the "spin
current" model is able to explain at least semi-quantitatively many
experimental data. This model was also applied to the systems with e$_{g}$
orbitals such as TbMnO$_{3}$[22]. In our opinion, the foremost task is the
embodiment of this idea in a crystal in which completely different properties
and behaviors can emerge from complexities and interrelations between
electrons, spins, and lattice structure. Equally important\ is the calculation
of the magnitude of electric polarization induced by "spin current". According
to KNB, the polarization is of the orders $eI(t/\Delta)$ or $eI(t/\Delta)^{3}$
for one or two holes. Here $I$ is the expectation value of length, $t$ the
hybridization energy and $\Delta$ the energy difference between d-orbitals and
p-orbitals. According to their estimation, $t=V(pd\pi)\sim0.1eV$,
$\Delta\approx2eV$, and $eI/a^{3}\sim10^{4}\mu C/m^{2}$ where $a$ is the
lattice constant. One can see that the magnitude is compatible with
experimental data only in very favorable conditions. Therefore, it is
desirable to conceive possible and realistic mechanism to enhance the
ferroelectricity-magnetism coupling.

KNB derived an elegant expression for the electric polarization of a
three-atom system $\overrightarrow{P}\sim eI\widehat{e}_{12}\times(\widehat
{e}_{1}\times\widehat{e}_{2})$ where $\widehat{e}_{1}$ and $\widehat{e}_{2}$
are the directions of spins of the transition metal ions and $\widehat{e}%
_{12}$ is the bond direction. We shall see how the expression conforms in a
crystal and how it is related to the wave vector of the helical magnetic
order. We will explain the reason why it is advantageous to have the helical
spin configuration for ferroelectricity and also elaborate the roles played by
the spin-orbit interaction. In fact, we will show that the electric
polarization comes directly from DM interaction. Finally, we point out that a
common feature in transition metal oxides, the bond-bending, can enhance
electric polarization.

\subsubsection{2. Description of the system}

We are going to consider two features in the system, the helical spin
configuration and bond-bending. Mostovoy[23] studied a system of degenerate
double-exchange interaction and next-nearest neighbor hopping. He found that
under certain conditions the helical spin configuration is stable. Thus, we
shall take that as our starting point. The transition metal ion, with position
vector $\overrightarrow{R}_{j,m}=\overrightarrow{R}_{j}+$ $\overrightarrow
{r}_{m}$ where $\overrightarrow{R}_{j}$ is the position vector of the j-th
lattice point and $\overrightarrow{r}_{m}$ is the position vector of the m-th
ion in the basis, has the following form for its spin (presumably those of
t$_{2g}$ electrons)
\begin{equation}
\overrightarrow{S}_{j,m}=S[(\widehat{e}_{x}\cos\phi+\widehat{e}_{y}\sin
\phi)\sin(\overrightarrow{q}\cdot\overrightarrow{R}_{j,m})+\widehat{e}_{z}%
\cos(\overrightarrow{q}\cdot\overrightarrow{R}_{j,m})] \tag{1}%
\end{equation}
The helical spin order has a wave vector $\overrightarrow{q}$. As shown in
Fig. 1, the projection of spins on xy-plane makes a fixed angle $\phi$ with x-axis.

Hund's coupling in the transition metal ions, $-J_{H}\Sigma\overrightarrow
{S}_{j,m}\cdot\overrightarrow{s}_{j,m}$, is the dominant mechanism in the
system. If the local spins in eq. (1) are treated as classical spins, their
effect on the spins of hybridizing electrons, denoted by $\overrightarrow
{s}_{j,m}$, is equivalent to an effective magnetic field. Thus, the eigen
state of the spin of a hybridizing electron on the m-th transition metal
ion\ with lower energy is
\begin{equation}
{\Big (}%
\begin{array}
[c]{c}%
\cos\theta_{j,m}\\
e^{i\phi}\sin\theta_{j,m}%
\end{array}
{\Big )} \tag{2}%
\end{equation}
where $\theta_{j,m}=\overrightarrow{q}\cdot\overrightarrow{R}_{j,m}/2$.\ This
will affect the hybridization of orbitals.

Next, we consider the hybridization. KNB studies a TM-O-TM\ three-atom triad.
Due to the symmetries of orbitals, the hybridization arises from $\pi
$-bonding. If the TM-O-TM triad is not linear (illustrated in Fig. 2a where
$\alpha$ is the bond angle), the $p_{x}$ orbitals of oxygen (if the bond is
approximately in the x-direction) can also take part in electron transfer,
$\sigma$-bonding can be realized and the hybridization energy can be greater.
In fact, the bond-bending occurs quite often in transition metal oxides. See,
for example, reference 24. The $ab$ plane of the crystal is shown in Fig. 2b
where the solid dots and circles denote the transition metal ions and oxygen
atoms respectively. There are two distinct transition metal ions in the basis,
and thus $m=1,2$, due to bond-bending. As a result, two transition metal ions
and four oxygen atoms form the basis of the crystal which is enlarged by
bond-bending. The atoms are labeled so as to facilitate later deduction. As
mentioned above, bond-bending will affect the hybridization energies. They can
be found in Slater and Koster[25].

Bond-bending has other profound effects. It changes the symmetry of the
surrounding of atoms and thus the symmetry of crystal field. In an
orthorhombic crystal $(a\neq b)$ with bond-bending, the $xy$-orbitals of the
transition metal ion will mix with the $x^{2}-y^{2}$ orbitals. The resulting
orbital has the form
\begin{equation}
\cos\beta|xy\rangle\mp\sin\beta|x^{2}-y^{2}\rangle, \tag{3}%
\end{equation}
where the $\pm$ sign is determined by the direction of the displacement of the
oxygen atom away from the line joining two transition metal ions. $\beta$ and
$\alpha$\ are of the same order of magnitudes. Ideally, if the crystal field
is determined by the four nearest oxygen ions on the $ab$ plane, then
$\beta=\alpha^{\prime}+\delta\alpha^{\prime}$ where $\alpha^{\prime}%
=\pi-\alpha$. The deviation $\delta\alpha^{\prime}$\ is due to the fact that
the symmetry of crystal field is not determined by bond-bending alone. We
expect $\delta\alpha^{\prime}$\ to be finite but smaller than $\alpha$ in
orthorhombic crystals of manganites but large in compounds like Ni$_{3}$%
V$_{2}$O$_{8}$. However, to get its magnitude, one need to carry out a very
precise first-principle calculation which is beyond the scope of the current paper.

The hybridization of the p-orbital of oxygen atoms and the xy-orbitals of
transition ions is equal to $\pm(\sqrt{3}/2)\sin\alpha\sin(\alpha
/2)V(pd\sigma)$ where the sign is again determined by the direction of the
displacements of oxygen atoms. That of the p-orbital of oxygen atoms and the
x$^{2}$-y$^{2}$-orbitals of transition ions is equal to $-(\sqrt{3}%
/2)\cos\alpha\sin(\alpha/2)V(pd\sigma)$. Although the x$^{2}$-y$^{2}$-orbitals
have greater hybridization energy, its amplitude is smaller as one can see
from (3). The resulting hybridization energy is $\pm(\sqrt{3}/2)\sin
(\alpha^{\prime}-\beta)\sin(\alpha/2)V(pd\sigma)$. We note in passing that
since $\alpha$ is close to $\pi$, if $|\beta-\alpha^{\prime}|\approx\pi
/12$,\ then $(\sqrt{3}/2)\sin(\alpha^{\prime}-\beta)\sin(\alpha/2)V(pd\sigma
)\approx0.2V(pd\sigma)$, which is often greater than $V(pd\pi)$.

\subsubsection{3. Calculation}

Due to the Hund's coupling, the spins of the hybridizing electron are
projected to the local spins and the hybridization energy is modified. It is
the main part of our Hamiltonian. We considered $\sigma$\ bond only. Hence,
the orbitals involved are that in (3) of transition metal ions, $p_{x}%
-$orbitals of oxygen atom 3 and 4 and\ $p_{y}-$orbitals of oxygen atom 5 and 6
where the numerals are shown in Fig. 2b. Later, we take the spin-orbit
interaction, and more orbitals, into account. It will be treated as a
perturbation. Our Hamiltonian thus has two parts, $H=H_{0}+H_{1}$:%

\begin{equation}
H_{0}=\sum\varepsilon_{p}c_{pil,\sigma}^{\dagger}c_{pil,\sigma}+\sum
\varepsilon_{d}c_{djm}^{\dagger}c_{djm}-\sum_{n.n.}V(-1)^{l}[\cos\theta
_{j,m}c_{djm}^{\dagger}c_{pil,\uparrow}+e^{i\phi}\sin\theta_{j,m}%
c_{djm}^{\dagger}c_{pil,\downarrow}]+H.c. \tag{4a}%
\end{equation}
and%
\begin{equation}
H_{1}=\lambda%
{\displaystyle\sum\limits_{j,m}}
\overrightarrow{l}_{j,m}\cdot\overrightarrow{s}_{j,m} \tag{4b}%
\end{equation}
where the hybridizations energy $V=(\sqrt{3}/2)\sin(\beta-\alpha^{\prime}%
)\sin(\alpha/2)V(pd\sigma)$, is only among nearest neighbors. $\lambda$ is the
strength of spin-orbit coupling. $\varepsilon_{p}$, $\varepsilon_{d}$ and
$c_{pil,\uparrow(\downarrow)}$, $c_{djm}$\ are the energies and field
operators of the $p-$orbitals in oxygen atoms and $d-$orbitals in transition
metal ions. $l$\ and $m$\ are the indices of oxygen atoms and transition metal
ions in the basis. As shown in Fig. 3, $m=1,2$\ for manganese ions and
$l=3-6$\ for oxygen atoms.\ As previously mentioned, the localized spins were
approximated as an effective magnetic field of angles\ $\theta_{j,m}$ and
$\phi$. We now made the following transformation
\begin{align}
p_{il,\uparrow}  &  =\cos(\overrightarrow{q}\cdot\overrightarrow{R}%
_{i,l}/2)c_{pil,\uparrow}+e^{i\phi}\sin(\overrightarrow{q}\cdot\overrightarrow
{R}_{i,l}/2)c_{pil,\downarrow}\tag{5a}\\
p_{il,\downarrow}  &  =\cos(\overrightarrow{q}\cdot\overrightarrow{R}%
_{i,l}/2)c_{pil,\downarrow}-e^{-i\phi}\sin(\overrightarrow{q}\cdot
\overrightarrow{R}_{i,l}/2)c_{pil,\uparrow} \tag{5b}%
\end{align}
where $\overrightarrow{R}_{i,l}=\overrightarrow{R}_{i}+$ $\overrightarrow
{r}_{l}$ with $\overrightarrow{r}_{l}$\ being the position vector of the
$l$-th oxygen atom in the basis without bond-bending. The Hamiltonian becomes
\begin{align}
H_{0}  &  =\sum\varepsilon_{p}p_{il,\sigma}^{\dagger}p_{il,\sigma}%
+\sum\varepsilon_{d}d_{jm}^{\dagger}d_{jm}-%
{\displaystyle\sum_{n.n.}}
V[(-1)^{l}\cos\delta\theta_{l}d_{jm}^{\dagger}p_{il,\uparrow}\nonumber\\
&  -\eta_{l}(-1)^{m}e^{i\phi}\sin\delta\theta_{l}d_{jm}^{\dagger
}p_{il,\downarrow}]+H.c. \tag{6}%
\end{align}
where $d_{jm}=c_{djm}$, $\delta\theta_{l}=a_{0}q_{x}/2$, $\eta_{l}=1$ for
$l=3,4$ and $\delta\theta_{l}=b_{0}q_{y}/2$, $\eta_{l}=-1$ for $l=5,6$.
$a_{0}$\ and $b_{0}$\ are,\ respectively,\ the distance between transition
metal ions along x and y direction. The Hamiltonian in momentum space is
\begin{align}
H_{0}  &  =\sum\varepsilon_{p}p_{lk,\sigma}^{\dagger}p_{lk,\sigma}%
+\sum\varepsilon_{d}d_{mk}^{\dagger}d_{mk}-\sum_{n.n.}V[(-1)^{l}\cos
\delta\theta_{l}e^{i\overrightarrow{k}\cdot(\overrightarrow{R}_{i,l}%
-\overrightarrow{R}_{j,m})}d_{mk}^{\dagger}p_{lk,\uparrow}\nonumber\\
&  -\eta_{l}(-1)^{m}e^{i\phi}\sin\delta\theta_{l}e^{i\overrightarrow{k}%
\cdot(\overrightarrow{R}_{i,l}-\overrightarrow{R}_{j,m})}d_{mk}^{\dagger
}p_{lk,\downarrow}]+H.c.. \tag{7}%
\end{align}
The following is its matrix form:
\begin{equation}
\left(
\begin{array}
[c]{cccccccccc}%
\varepsilon_{d} & 0 & V_{+x}c_{3} & -V_{+x}e^{i\phi}s_{3} & -V_{-x}c_{4} &
-V_{-x}e^{i\phi}s_{4} & V_{-y}c_{5} & V_{-y}e^{i\phi}s_{5} & -V_{+y}c_{6} &
V_{+y}e^{i\phi}s_{6}\\
0 & \varepsilon_{d} & V_{-x}c_{3} & V_{-x}e^{+i\phi}s_{3} & -V_{+x}c_{4} &
V_{+x}e^{i\phi}s_{4} & V_{+y}c_{5} & -V_{+y}e^{i\phi}s_{5} & -V_{-y}c_{6} &
-V_{-y}e^{i\phi}s_{6}\\
V_{-x}c_{3} & V_{+x}c_{3} & \varepsilon_{p} & 0 & 0 & 0 & 0 & 0 & 0 & 0\\
-V_{-x}e^{-i\phi}s_{3} & V_{+x}e^{-i\phi}s_{3} & 0 & \varepsilon_{p} & 0 & 0 &
0 & 0 & 0 & 0\\
-V_{+x}c_{4} & -V_{-x}c_{4} & 0 & 0 & \varepsilon_{p} & 0 & 0 & 0 & 0 & 0\\
-V_{+x}e^{-i\phi}s_{4} & V_{-x}e^{-i\phi}s_{4} & 0 & 0 & 0 & \varepsilon_{p} &
0 & 0 & 0 & 0\\
V_{+y}c_{5} & V_{-y}c_{5} & 0 & 0 & 0 & 0 & \varepsilon_{p} & 0 & 0 & 0\\
V_{+y}e^{-i\phi}s_{5} & -V_{-y}e^{-i\phi}s_{5} & 0 & 0 & 0 & 0 & 0 &
\varepsilon_{p} & 0 & 0\\
-V_{-y}c_{6} & -V_{+y}c_{6} & 0 & 0 & 0 & 0 & 0 & 0 & \varepsilon_{p} & 0\\
V_{-y}e^{-i\phi}s_{6} & -V_{+y}e^{-i\phi}s_{6} & 0 & 0 & 0 & 0 & 0 & 0 & 0 &
\varepsilon_{p}%
\end{array}
\right)  \tag{8}%
\end{equation}
Here the order of matrix elements is $m=1,2$, and then $l=3\uparrow
,3\downarrow,4\uparrow,4\downarrow$ etc.., $V_{\pm x}=Ve^{\pm ik_{x}a_{0}}%
$,\ $V_{\pm y}=Ve^{\pm ik_{y}b_{0}}$\ and\ $s_{i}=\sin\delta\theta_{i}$\ ,
$c_{i}=\cos\delta\theta_{i}$. It can be diagonalized and the eigen values are
\begin{equation}
E_{1\pm}=\frac{\varepsilon_{d}+\varepsilon_{p}}{2}+\sqrt{(\frac{\varepsilon
_{d}-\varepsilon_{p}}{2})^{2}+4V^{2}\pm V^{2}{\big |}%
{\displaystyle\sum\limits_{n.n.}}
\cos\delta\theta_{nn}e^{2i\overrightarrow{k}\cdot\overrightarrow{R}_{nn}%
}{\big |}} \tag{9a}%
\end{equation}%
\begin{equation}
E_{2\pm}=\frac{\varepsilon_{d}+\varepsilon_{p}}{2}-\sqrt{(\frac{\varepsilon
_{d}-\varepsilon_{p}}{2})^{2}+4V^{2}\pm V^{2}{\big |}%
{\displaystyle\sum\limits_{n.n.}}
\cos\delta\theta_{nn}e^{2i\overrightarrow{k}\cdot\overrightarrow{R}_{nn}%
}{\big |}} \tag{9b}%
\end{equation}

\begin{equation}
E_{3}=\varepsilon_{p} \tag{9c}%
\end{equation}
where $\varepsilon_{p}$ is six-fold degenerate and $\overrightarrow{R}_{nn}$
and $\delta\theta_{nn}=2\delta\theta_{l}$ are the position vector and spin
angle difference between two nearest neighboring transition element ions. The
eigen vectors are%
\begin{equation}
\psi_{1+,\overrightarrow{k}\boldsymbol{,}\overrightarrow{q}}(\overrightarrow
{r})=%
{\displaystyle\sum_{j,m}}
e^{i\overrightarrow{k}\cdot\overrightarrow{R}_{j,m}}A_{m1+}{\Big [}\psi
_{d}(\overrightarrow{r}-\overrightarrow{R}_{j,m})+%
{\displaystyle\sum\limits_{l}}
\frac{V(-1)^{l}e^{i\overrightarrow{k}\cdot\overrightarrow{r}_{lm}}}%
{E_{1+}-\varepsilon_{p}}\psi_{pl}(\overrightarrow{r}-\overrightarrow{R}%
_{j,l}){\Big ](}%
\begin{array}
[c]{c}%
\cos\theta_{j,m}\\
e^{i\phi}\sin\theta_{j,m}%
\end{array}
{\Big )} \tag{10a}%
\end{equation}%
\begin{equation}
\psi_{1-,\overrightarrow{k}\boldsymbol{,}\overrightarrow{q}}(\overrightarrow
{r})=%
{\displaystyle\sum_{j,m}}
e^{i\overrightarrow{k}\cdot\overrightarrow{R}_{j,m}}A_{m1-}{\Big [}\psi
_{d}(\overrightarrow{r}\psi_{pl})+%
{\displaystyle\sum\limits_{l}}
\frac{V(-1)^{l}e^{i\overrightarrow{k}\cdot\overrightarrow{r}_{lm}}}%
{E_{1-}-\varepsilon_{p}}\psi_{pl}(\overrightarrow{r}-\overrightarrow{R}%
_{j,l}){\Big ](}%
\begin{array}
[c]{c}%
\cos\theta_{j,m}\\
e^{i\phi}\sin\theta_{j,m}%
\end{array}
{\Big )} \tag{10b}%
\end{equation}%
\begin{equation}
\psi_{2+,\overrightarrow{k}\boldsymbol{,}\overrightarrow{q}}(\overrightarrow
{r})=%
{\displaystyle\sum_{j,m}}
e^{i\overrightarrow{k}\cdot\overrightarrow{R}_{j,m}}A_{m2+}{\Big [}\psi
_{d}(\overrightarrow{r}-\overrightarrow{R}_{j,m})+%
{\displaystyle\sum\limits_{l}}
\frac{V(-1)^{l}e^{i\overrightarrow{k}\cdot\overrightarrow{r}_{lm}}}%
{E_{2+}-\varepsilon_{p}}\psi_{pl}(\overrightarrow{r}-\overrightarrow{R}%
_{j,l}){\Big ](}%
\begin{array}
[c]{c}%
\cos\theta_{j,m}\\
e^{i\phi}\sin\theta_{j,m}%
\end{array}
{\Big )} \tag{10c}%
\end{equation}%
\begin{equation}
\psi_{2-,\overrightarrow{k}\boldsymbol{,}\overrightarrow{q}}(\overrightarrow
{r})=%
{\displaystyle\sum_{j,m}}
e^{i\overrightarrow{k}\cdot\overrightarrow{R}_{j,m}}A_{m2-}{\Big [}\psi
_{d}(\overrightarrow{r}-\overrightarrow{R}_{j,m})+%
{\displaystyle\sum\limits_{l}}
\frac{V(-1)^{l}e^{i\overrightarrow{k}\cdot\overrightarrow{r}_{lm}}}%
{E_{2-}-\varepsilon_{p}}\psi_{pl}(\overrightarrow{r}-\overrightarrow{R}%
_{j,l}){\Big ](}%
\begin{array}
[c]{c}%
\cos\theta_{j,m}\\
e^{i\phi}\sin\theta_{j,m}%
\end{array}
{\Big )} \tag{10d}%
\end{equation}
where $A_{m1(2)\pm}$ are normalization constants, $\overrightarrow{r}%
_{lm}=\overrightarrow{r}_{l}-\overrightarrow{r}_{m}$, and
\begin{equation}
\frac{A_{m1\pm}}{A_{m2\pm}}=\pm\frac{%
{\displaystyle\sum\limits_{n.n.}}
\cos\delta\theta_{nn}e^{2i\overrightarrow{k}\cdot\overrightarrow{R}_{nn}}%
}{{\big |}%
{\displaystyle\sum\limits_{n.n.}}
\cos\delta\theta_{nn}e^{2i\overrightarrow{k}\cdot\overrightarrow{R}_{nn}%
}{\big |}}. \tag{11}%
\end{equation}
Note that $\psi_{pl}=\psi_{px}$\ for $l=3,4$\ and $\psi_{pl}=\psi_{py}$\ for
$l=5,6$. The eigen vectors of $\varepsilon_{p}$ do not concern us because they
are non-bonding states. If there is no bond-bending, then there are only $\pi
$-bonding for xy-orbitals. It turns out that the eigen vectors have very
similar forms as those in eqs. (9). They are shown in Appendix.

\subsubsection{4. Spin-orbit interaction and polarization}

We now introduce the spin-orbit interaction in eq. (4). Its effect can be
expressed in the following relations%

\begin{equation}
\overrightarrow{l}\cdot\overrightarrow{s}\,|xy,\uparrow\rangle=i|zx,\downarrow
\rangle+|yz,\downarrow\rangle-i|x^{2}-y^{2},\uparrow\rangle\tag{12a}%
\end{equation}

\begin{equation}
\overrightarrow{l}\cdot\overrightarrow{s}\,|xy,\downarrow\rangle
=i|zx,\uparrow\rangle-|yz,\uparrow\rangle+i|x^{2}-y^{2},\downarrow
\rangle\tag{12b}%
\end{equation}%
\begin{equation}
\overrightarrow{l}\cdot\overrightarrow{s}\,|x^{2}-y^{2},\uparrow
\rangle=|zx,\downarrow\rangle-i|yz,\downarrow\rangle+i|xy,\uparrow
\rangle\tag{12c}%
\end{equation}%
\begin{equation}
\overrightarrow{l}\cdot\overrightarrow{s}\,|x^{2}-y^{2},\downarrow
\rangle=-|zx,\uparrow\rangle-i|yz,\uparrow\rangle-i|xy,\downarrow\rangle.
\tag{12d}%
\end{equation}
Therefore, if the spin-orbit interaction is treated perturbatively, the wave
functions $\psi_{d}(\overrightarrow{r}-\overrightarrow{R}_{j,m})(\cos
\theta_{j,m}|\uparrow\rangle+e^{i\phi}\sin\theta_{j,m}|\downarrow\rangle)$ in
eqs. (10) will be replaced by%
\begin{align}
&  -\cos\beta{\Big \{}\psi_{xy}(\overrightarrow{r}-\overrightarrow{R}%
_{j,m}){\Big (}%
\begin{array}
[c]{c}%
\cos\theta_{j,m}\\
e^{i\phi}\sin\theta_{j,m}%
\end{array}
{\Big )}+\frac{\lambda}{\Delta E_{cf}}{\Big [}\psi_{x^{2}-y^{2}}%
(\overrightarrow{r}-\overrightarrow{R}_{j,m}){\Big (}%
\begin{array}
[c]{c}%
-i\cos\theta_{j,m}\\
ie^{i\phi}\sin\theta_{j,m}%
\end{array}
{\Big )}\nonumber\\
&  +\psi_{zx}(\overrightarrow{r}-\overrightarrow{R}_{j,m}){\Big (}%
\begin{array}
[c]{c}%
ie^{i\phi}\sin\theta_{j,m}\\
i\cos\theta_{j,m}%
\end{array}
{\Big )}+\psi_{yz}(\overrightarrow{r}-\overrightarrow{R}_{j,m}){\Big (}%
\begin{array}
[c]{c}%
-e^{i\phi}\sin\theta_{j,m}\\
\cos\theta_{j,m}%
\end{array}
{\Big )]\}}\nonumber\\
&  \pm\sin\beta{\Big \{}\psi_{x^{2}-y^{2}}(\overrightarrow{r}-\overrightarrow
{R}_{j,m}){\Big (}%
\begin{array}
[c]{c}%
\cos\theta_{j,m}\\
e^{i\phi}\sin\theta_{j,m}%
\end{array}
{\Big )}+\frac{\lambda}{\Delta E_{cf}}{\Big [}\psi_{xy}(\overrightarrow
{r}-\overrightarrow{R}_{j,m}){\Big (}%
\begin{array}
[c]{c}%
i\cos\theta_{j,m}\\
-ie^{i\phi}\sin\theta_{j,m}%
\end{array}
{\Big )}\nonumber\\
&  +\psi_{zx}(\overrightarrow{r}-\overrightarrow{R}_{j,m}){\Big (}%
\begin{array}
[c]{c}%
-e^{i\phi}\sin\theta_{j,m}\\
\cos\theta_{j,m}%
\end{array}
{\Big )}+\psi_{yz}(\overrightarrow{r}-\overrightarrow{R}_{j,m}){\Big (}%
\begin{array}
[c]{c}%
-ie^{i\phi}\sin\theta_{j,m}\\
-i\cos\theta_{j,m}%
\end{array}
{\Big )]\}} \tag{13}%
\end{align}
where $\Delta E_{cf}$ is the energy difference between $\cos\beta|xy\rangle
\mp\sin\beta|x^{2}-y^{2}\rangle$ in (2) and $\sin\beta|xy\rangle\pm\cos
\beta|x^{2}-y^{2}\rangle$. Substituting (13) into eqs. (10), we found that the
polarization per unit cell is
\begin{align}
\overrightarrow{P}_{\overrightarrow{q}}  &  =\left\langle -e\overrightarrow
{r}\right\rangle =%
{\displaystyle\sum\limits_{\boldsymbol{k}}}
{\displaystyle\int}
d^{3}\overrightarrow{r}\,\psi_{\overrightarrow{k}\boldsymbol{,}\overrightarrow
{q}}^{\ast}(\overrightarrow{r})(-e\overrightarrow{r})\psi_{\overrightarrow
{k}\boldsymbol{,}\overrightarrow{q}}(\overrightarrow{r})\nonumber\\
&  \approx\frac{8\sqrt{2}\sin\beta V\lambda}{\pi(\varepsilon_{p}%
-\varepsilon_{d})\Delta E_{cf}}(e\mathbb{\rho}\widehat{e}_{z})[\cos\phi
\sin(a_{0}q_{x}/2)+\sin\phi\sin(b_{0}q_{y}/2)] \tag{14}%
\end{align}
where
\begin{equation}
\rho\approx-%
{\displaystyle\int}
d^{3}\overrightarrow{r}\,\psi_{zx}^{\ast}(\overrightarrow{r})\smallskip
z\smallskip\psi_{px}(\overrightarrow{r}-a_{0}\widehat{e}_{x}/2)\approx-%
{\displaystyle\int}
d^{3}\overrightarrow{r}\,\psi_{yz}^{\ast}(\overrightarrow{r})\smallskip
z\smallskip\psi_{py}(\overrightarrow{r}-b_{0}\widehat{e}_{y}/2) \tag{15}%
\end{equation}
with $a_{0}$\ and $b_{0}$\ being the lattice constants without bond-bending.
Another way of writing them should be $\overrightarrow{a}_{0}=(\overrightarrow
{R}_{j+1}-\overrightarrow{R}_{j})/2$ and $\overrightarrow{b}_{0}%
=(\overrightarrow{R}_{j+1}-\overrightarrow{R}_{j})/2$\ for site $j+1$%
\ situated at x- or y-direction away from site $j$.\ There are ten bands in
our calculation. The main contribution comes from the topmost occupied band.
We have also made the approximation $E_{n+}\approx\varepsilon_{p}$, taking the
advantage of the fact that $V\ll|\varepsilon_{p}-\varepsilon_{d}|$.\ If even
number of bands are occupied, the polarization will be much smaller. The
polarization produced by two bands tends to cancel each other. As a result,
there is an extra factor of $4V^{2}/(\varepsilon_{p}-\varepsilon_{d})^{2}$
which comes from the denominator of eqs. (10):
\begin{equation}
\overrightarrow{P}_{\overrightarrow{q}}\approx\frac{32\sqrt{2}\sin\beta
V^{3}\lambda}{\pi(\varepsilon_{p}-\varepsilon_{d})^{3}\Delta E_{cf}%
}(e\mathbb{\rho}\widehat{e}_{z})[\cos\phi\sin(a_{0}q_{x}/2)+\sin\phi\sin
(b_{0}q_{y}/2)] \tag{16}%
\end{equation}

To see more clearly how polarization and $\overrightarrow{q}$ are related, let
$U_{P}$ be the space inversion operator. We then have%

\begin{equation}
U_{P}\psi_{\overrightarrow{k}\boldsymbol{,}\overrightarrow{q}}(\overrightarrow
{r})=\psi_{-\overrightarrow{k}\boldsymbol{,-}\overrightarrow{q}}%
(\overrightarrow{r}). \tag{17}%
\end{equation}
Under the space inversion, the displacements of oxygen atoms and hence,
$\alpha$,\ $\beta$\ and $V$\ change sign under inversion
\begin{equation}
\overrightarrow{P}_{\overrightarrow{q}}=-e\left\langle U_{P}U_{P}%
\overrightarrow{r}U_{P}U_{P}\right\rangle =-\overrightarrow{P}%
_{-\overrightarrow{q}} \tag{18}%
\end{equation}
and the polarization is an odd function of $\overrightarrow{q}$. The form of
sine function in eqs. (14) and (16) seems to be a natural form. In the
continuum limit ($a_{0}\approx b_{0}\longrightarrow0$) or the long wavelength
limit ($\overrightarrow{q}\longrightarrow0$), we found
\begin{equation}
\overrightarrow{P}_{\overrightarrow{q}}\sim0.01e\mathbb{\rho}a_{0}\widehat
{e}_{z}(q_{x}\cos\phi+q_{y}\sin\phi)\sim0.01e\mathbb{\rho}a_{0}\overrightarrow
{q}\times\widehat{h} \tag{19}%
\end{equation}
where $\widehat{h}$\ is the helix axis unit vector. We have assumed that
$q_{z}=0$ since our calculation was performed on xy-plane. If one recognizes
that the true meaning of $\widehat{e}_{12}$\ is the spatial direction alone
which the spins propagate, i.e., $\overrightarrow{q}$,\ he can see how the
form $\overrightarrow{P}\sim eI\widehat{e}_{12}\times(\widehat{e}_{1}%
\times\widehat{e}_{2})$ given by KNB\ can be transformed in the presence of
magnetic orders.\ 

\subsubsection{5. Discussion}

In order to see how magnetic orders and electric order are coupled, let us go
back to the original work of Moriya[19]. He derived the following expression:
\begin{equation}
H_{DM}=%
{\displaystyle\sum\limits_{N}}
\overrightarrow{D}_{N,N-1}\cdot(\overrightarrow{S}_{N}\times\overrightarrow
{S}_{N-1}) \tag{20}%
\end{equation}
where
\begin{equation}
\overrightarrow{D}_{N,N-1}=i\lambda%
{\displaystyle\sum}
\frac{J(n,n^{\prime},m,n^{\prime})\left\langle n|\overrightarrow{l}%
_{i}|m\right\rangle }{E_{n}-E_{m}}-i\lambda%
{\displaystyle\sum}
\frac{J(n,n^{\prime},n,m^{\prime})\left\langle n^{\prime}|\overrightarrow
{l}_{j}|m^{\prime}\right\rangle }{E_{n^{\prime}}-E_{m^{\prime}}}. \tag{21}%
\end{equation}
Here, $J(n,n^{\prime},m,m^{\prime})$ is the exchange interaction strength and
$\overrightarrow{l}_{i}$ denotes the angular momentum of the electron at site
$i$.

In our starting Hamiltonian, the exchange interaction comes from the charge
transfer energy $\varepsilon_{p}-\varepsilon_{d}$\ and hybridization energy
$V$[27]. Combined with the spin-orbit interaction in eq. (4b), DM interaction
is clearly present in the system we considered. We can recast the wave
functions we got previously in a form similar to Moriya's by treating the
hybridization energy and spin-orbit interaction as perturbations,%
\begin{equation}
\overrightarrow{P}_{\overrightarrow{q}}=-e%
{\displaystyle\sum_{n,m,l}}
\frac{\left\langle m|H_{V}|n\right\rangle \lambda\left\langle
n|\overrightarrow{l}|l\right\rangle \cdot\left\langle \sigma_{n}%
|\overrightarrow{s}|\sigma_{l}\right\rangle }{(\varepsilon_{d}-\varepsilon
_{p})\Delta E_{cf}}\left\langle l,\sigma_{l}|\overrightarrow{r}|m,\sigma
_{m}\right\rangle +c.c. \tag{22}%
\end{equation}
where $H_{V}$ is the hybridization energy (the third and fourth terms of eq.
(4a)).\ $|l\rangle$\ and $|\sigma_{l}\rangle$\ are respectively the spatial
and spin part of an intermediate state.\ $|n\rangle$\ and $|l\rangle$\ are
states of d-orbitals and $|m\rangle$\ p-orbitals.\ $\left\langle
n|H_{V}|m\right\rangle /(\varepsilon_{d}-\varepsilon_{p})$\ is the exchange
part with $H_{V}$\ also containing information of spins. $\lambda\left\langle
l|\overrightarrow{l}|n\right\rangle \cdot\left\langle \sigma_{l}%
|\overrightarrow{s}|\sigma_{n}\right\rangle $\ is the spin-orbit coupling. Eq.
(22) is also applicable to the situations without bond-bending. In that case,
the $\pi$-bonding will be considered as they were in KNB's original work and
$H_{V}$ is the third term of eq. (A-1). One can see that eq. (22) has the same
origin as $H_{DM}$\ in eqs. (20) and (21). However, we have to note that the
helical spin configuration is not caused by DM interaction whose strength is
too small. Rather, it can be due to the next-near-neighbor hybridization as
shown in ref. 23.

It is easier to analyze with the following form:
\begin{equation}
P_{\overrightarrow{q},k}=-e%
{\displaystyle\sum_{n,m,l,j}}
\frac{\left\langle m|H_{V}|n\right\rangle \lambda\left\langle n|\varepsilon
_{hij}r_{h}p_{i}|l\right\rangle \left\langle \sigma_{n}|s_{j}|\sigma
_{l}\right\rangle \left\langle l,\sigma_{l}|r_{k}|m,\sigma_{m}\right\rangle
}{(\varepsilon_{d}-\varepsilon_{p})\Delta E_{cf}}+c.c.. \tag{23}%
\end{equation}
where $\varepsilon_{hij}$\ is the antisymmetric Levi-Cevita symbol and $r_{k}%
$\ is the $k$-th component\ of $\overrightarrow{r}$\ in space. We consider the
mirror symmetry of above equation. $H_{V}$ may change sign under mirror
reflection operation because of the orbital wave functions involved. It is
also related to the direction of the displacements of oxygen atoms. For
example, the (pd$\pi$) part of $E_{y,xy}$ and (pd$\sigma$)\ of $E_{x,x^{2}%
-y^{2}}$\ ( in Slater-Koster notation) with bond along x-direction change sign
if one makes the operation $x\longrightarrow-x$. On the other hand, the
(pd$\sigma$) part of $E_{x,xy}$\ (due to bond-bending) does not change sign if
one makes the operation $x\longrightarrow-x$\ or $y\longrightarrow-y$.\ In the
previous section, we calculated the polarization of a planar crystal. Now we
consider a more general case of orthorhombic structure and helical spin
configuration. It is easier to catch the essence if one considers the (pd$\pi
$) part of $E_{y,xy}$ or the (pd$\sigma$)\ part of $E_{x,x^{2}-y^{2}}$ of
$H_{V}$, which is applicable to KNB's original work and to the $x^{2}-y^{2}%
$\ orbital part of our work respectively.\ Assuming the bond direction of
$H_{V}$\ is in the $m$-direction then polarization is finite for either $k=h$,
$m=i$\ or $k=i$, $m=h$.\ As a result, eq. (23) can be simplified as
\begin{equation}
P_{\overrightarrow{q},k}=-2e%
{\displaystyle\sum_{n,m,l,j}}
\frac{\left\langle m|H_{V,i}|n\right\rangle \lambda\left\langle n|l_{j}%
|l\right\rangle \left\langle \sigma_{n}|s_{j}|\sigma_{l}\right\rangle
}{(\varepsilon_{d}-\varepsilon_{p})}\frac{\left\langle l,\sigma_{l}%
|r_{k}|m,\sigma_{m}\right\rangle }{\Delta E_{cf}}+c.c.. \tag{24}%
\end{equation}
where $H_{V,i}$\ denotes the hybridization bond along $i-$direction
and\ $i$,\ $j$\ and $k$\ are cyclic.

The spin part needs more attention. $\left\langle \sigma_{n}|s_{j}|\sigma
_{l}\right\rangle $\ give rise to a spin state $|\sigma_{l}\rangle$ different
from the original state $|\sigma_{n}\rangle$\ and\ $\left\langle \sigma
_{l}|\sigma_{m}\right\rangle $\ gives rise to a interesting
contribution.\ Only the imaginary part\ needs to be considered because
$\left\langle l|l_{j}|n\right\rangle $ is imaginary:
\begin{align}
\operatorname{Im}(\left\langle \sigma_{n}|s_{x}|\sigma_{l}\right\rangle
\left\langle \sigma_{l,x}|\sigma_{n}\right\rangle )  &  =\sin\phi
\sin(\overrightarrow{q}\cdot\overrightarrow{R}_{N,N-1}/2)=(\overrightarrow
{s}_{N}\times\overrightarrow{s}_{N-1})|_{x}\tag{25a}\\
\operatorname{Im}(\left\langle \sigma_{n}|s_{y}|\sigma_{l}\right\rangle
\left\langle \sigma_{l,y}|\sigma_{n}\right\rangle )  &  =-\cos\phi
\sin(\overrightarrow{q}\cdot\overrightarrow{R}_{N,N-1}/2)=(\overrightarrow
{s}_{N}\times\overrightarrow{s}_{N-1})|_{y} \tag{25b}%
\end{align}
for the spin configuration in eq. (1). Here the additional subscript $j$\ of
$|\sigma_{l,j}\rangle$\ denotes that it comes from $\left\langle \sigma
_{n}|s_{j}|\sigma_{l}\right\rangle $. The right hand sides of eqs. (24) are
very similar to respective components of $\overrightarrow{S}_{N}%
\times\overrightarrow{S}_{N-1}$ which appears in $H_{DM}$.\ The difference is
in the arguments of $\sin$ functions. The factor $1/2$\ arises because in our
model it is the hybridization electrons that mediate the exchange interaction
while in\ the original DM\ interaction\ it is direct exchange. Substituting
eqs. (25) into eq. (24),\ we found that
\begin{equation}
P_{\overrightarrow{q},k}=-2e%
{\displaystyle\sum_{n,m,l,j}}
\frac{\lambda\left\langle m|H_{V,i}|n\right\rangle }{(\varepsilon
_{d}-\varepsilon_{p})}\left\langle n|l_{j}|l\right\rangle (\overrightarrow
{s}_{N}\times\overrightarrow{s}_{N-1})|_{j}\frac{\left\langle l|r_{k}%
|m\right\rangle }{\Delta E_{cf}}+c.c.. \tag{26}%
\end{equation}

The connection between the electric polarization and spin current can now be
established. A common definition of the spin current is
\begin{equation}
\widetilde{j}_{i}^{j}=\frac{t}{4i\hbar}%
{\displaystyle\sum\limits_{N}}
(d_{\mathbf{R}_{N}+\mathbf{a}}^{+}\sigma_{j}d_{\mathbf{R}_{N}}-d_{\mathbf{R}%
_{N}}^{+}\sigma_{j}d_{\mathbf{R}_{N}+\mathbf{a}})\overrightarrow{a}|_{i}
\tag{27}%
\end{equation}
where $\overrightarrow{a}$ is the lattice vector of length $a_{0}$ in the
direction of \ $i$ and in our case $t\approx V^{2}/|\varepsilon_{dp}|$. Note
that the electrons hop along the direction of $\overrightarrow{a}$.\ Equation
(27) can be derived by discretizing the form $i\hbar\lbrack\psi^{+}\sigma
_{j}\partial_{i}\psi-(\partial_{i}\psi^{+})\sigma_{j}\psi]/4m$.\ The form in
eq. (27) manifests itself in helical spin configuration. We calculated its
expectation value with the state in (2) and found that\
\begin{equation}
\left\langle \widetilde{j}_{x}^{y}\right\rangle =\frac{ta_{0}}{2\hbar}\cos
\phi\sin(q_{x}a_{0}/2)=-\frac{ta_{0}}{2\hbar}(\overrightarrow{s}_{N}%
\times\overrightarrow{s}_{N-1})|_{y}, \tag{28a}%
\end{equation}%
\begin{equation}
\left\langle \widetilde{j}_{y}^{x}\right\rangle =-\frac{ta_{0}}{2\hbar}%
\sin\phi\sin(q_{y}a_{0}/2)=-\frac{ta_{0}}{2\hbar}(\overrightarrow{s}_{N}%
\times\overrightarrow{s}_{N-1})|_{x}. \tag{28b}%
\end{equation}
The results are same as those in eqs. (25). Introducing the SU(2) vector
potential[28] $\mathcal{A}_{j}^{i}=-\varepsilon_{ijk}\hbar E_{k}/4m_{e}c^{2}$
where $m_{e}\ $is\ the electron mass,\ the DM interaction can be rewritten as
\begin{equation}
H_{DM}=e%
{\displaystyle\sum}
\mathcal{A}_{j}^{i}\widetilde{j}_{i}^{j}. \tag{29}%
\end{equation}
where
\begin{equation}
\mathcal{A}_{j}^{i}=-\frac{2J\lambda}{eta_{0}\Delta E_{cf}}e_{B,i}l_{j},
\tag{30}%
\end{equation}
is the guage field coupled to the spin current. It comes from the term
$\overrightarrow{E}\cdot(\overrightarrow{p}\times\overrightarrow{\sigma})$ in
Pauli's equation. See for example, ref. 28. The polarization results from the
perturbation of $H_{DM}$:\
\begin{equation}
P_{\overrightarrow{q},k}=-e%
{\displaystyle\sum\limits_{M}}
\frac{\left\langle M|H_{DM}|0\right\rangle }{E_{0M}}\left\langle
0|r_{k}|M\right\rangle \tag{31}%
\end{equation}
where $|0\rangle$,\ the ground state, has a component $(V/\varepsilon
_{dp})|\psi_{p}\rangle$.\ If we take $E_{0M}$\ to be the exchange energy
$J$,\ then eq. (25) and eq. (30) are equivalent in view of eqs. (26) and (28-30).

Now we can see more clearly what the origin of internal electric field $E_{k}%
$\ is.\ From eq. (30), we found
\begin{equation}
\overrightarrow{E}=\frac{8m_{e}c^{2}J\lambda}{eta_{0}\Delta E_{cf}}\widehat
{e}_{B}\times\overrightarrow{l} \tag{32}%
\end{equation}
where the factor $m_{e}c^{2}$\ will be cancelled by its inverse in $\lambda
$.\ The small factor $J/t$\ arises because of the cancellation between
different band. See also eqs. (14) and (16). The electric field originates
from the perturbation of spin-orbit interaction.\ The spin-orbit interaction
changes the angular dependence of the d-orbital wave functions and hence, the
electron density, via the hybridization of d-and p-orbitals. This effect can
be interpreted as the result of an internal electric field. However, the
electric field thus created does not necessarily give rise to net electric
polarization. Certain environments are more advantageous than others. As one
can see from eqs. (20), (25) and (28), the helical spin configuration is apt
to provide spin current, net electric polarization and multiferroics.

The magnitude of polarization is also important. If there are odd number of
filled bands then the polarization is of the order $\overrightarrow
{P}_{\overrightarrow{q}}/\Omega$ where $\Omega\sim250\mathring{A}^{3}$\ is the
volume of a unit cell and eq. (14) is used. If we take $|\varepsilon
_{p}-\varepsilon_{d}|\sim2eV$, $\Delta E_{cf}\sim2.0eV$,\ $V=(\sqrt{3}%
/2)\sin(\alpha-\beta)\sin(\alpha/2)V(pd\sigma)\sim0.2eV$,\ $\lambda\sim
0.05eV$\ and $\sin\beta\sim0.1$\ then $P\sim10\mu C/m^{2}$\ for $\rho
\sim0.1\mathring{A}$. The bond-bending activated\ hybridization $V$\ is in
general greater than $V(pd\pi)$.\ and $\rho\sin\beta/I$ where $I$ is defined
in eq. (A-6), is of the order $V(pd\sigma)\sin\beta/V(pd\pi)\sim1$ for
$\beta\approx\pi/12$. Hence bond-bending gives larger polarization in many
oxides. For example, $\alpha\approx5\pi/6$ in manganites[26] and in compounds
such as Ni$_{3}$V$_{2}$O$_{8}$ the bond angle of Ni-O-Ni $\alpha$\ can be as
small as $\pi/2$[6]. Hence, $\alpha$\ and\ $\beta$\ may both be large and the
environment is favorable to ferroelectricity.\ Furthermore, $\rho
\sim0.1\mathring{A}$\ can very well be an underestimation because the $O^{2-}%
$\ has a much larger radius than a neutral oxygen atom. All things considered,
$P$\ can be an order of magnitude greater than previously estimated. Atomic
displacement can also enhance electric polarization by destroy cancellation.
However, if it has its own wave vector and it is not commensurate with
$\overrightarrow{q}$\ then there is no net polarization.

In conclusion, we have analyzed the conditions for the emergence of
ferroelectricity due to magnetic orders. We found a simple relation between
its wave vector $\overrightarrow{q}$\ and polarization. Furthermore, the
physical picture of the coupling between magnetic orders and ferroelectricity
is made clear. Multiferroics is created by a generalized version of
Dzyaloshinskii-Moriya interaction in the environment of certain spin orders,
preferably helical.\ Above findings can also be applied to systems without
bond-bending but with $\pi$-bond hybridization.\ The bond-bending tends to
enhance polarization and it may be important for certain compounds.

The author benefited from the activities of "quantum novel phenomena in
condensed matter" focus groups of NCTS, Taiwan and discussion with S. Maekawa.
This work is supported in part by the National Science Council under the
contract NSC 95-2112-M-002-048-MY3.

\subsubsection{Appendix}

In this Appendix, we present the eigen values and eigen vectors of the $\pi
$-bonding systems. It is simpler because there are only one transition element
ion (thus the dropping of the index $m$) and two oxygen atoms in the basis. We
consider a planar crystal on the xy-plane. For bonds along x(y)-direction,
p$_{y(x)}$-orbital of the oxygen atoms and the xy-orbital of the transition
element ions form $\pi$-bond. The p$_{z}$-orbitals are ignored because they do
not give rise to polarization in z-direction.orbitals are considered. Thus, we
have the Hamiltonian:
\begin{equation}
H_{0}=\sum\varepsilon_{p}c_{pil,\sigma}^{\dagger}c_{pil,\sigma}+\sum
\varepsilon_{d}c_{dj}^{\dagger}c_{dj}-\sum_{n.n.}V^{\prime}(-1)^{l}[\cos
\theta_{j}c_{dj}^{\dagger}c_{pil,\uparrow}+e^{i\phi}\sin\theta_{j}%
c_{dj}^{\dagger}c_{pil,\downarrow}]+H.c. \tag{A-1}%
\end{equation}
where $V^{\prime}=V(pd\pi)$ and $l=1,2$ for the oxygen atoms on x-axis and
y-axis respectively.\ Making a transformation similar to that in eqs (5), we
obtain the Hamiltonian in momentum space:
\begin{align}
H_{0}  &  =\sum\varepsilon_{p}p_{lk,\sigma}^{\dagger}p_{lk,\sigma}%
+\sum\varepsilon_{d}d_{k}^{\dagger}d_{k}-\sum_{n.n.}V^{\prime}(-1)^{l}%
[\cos\delta\theta_{l}e^{(\overrightarrow{R}_{i,l}-\overrightarrow{R}_{j}%
)}d_{k}^{\dagger}p_{lk,\uparrow}\nonumber\\
&  -e^{i\phi}\sin\delta\theta_{l}e^{i\overrightarrow{k}\cdot(\overrightarrow
{R}_{i,l}-\overrightarrow{R}_{j})}d_{k}^{\dagger}p_{lk,\uparrow}]+H.c..
\tag{A-2}%
\end{align}
Now $\delta\theta_{l=3,4}=q_{x}a_{0}/2$ and\ $\delta\theta_{l=5,6}=q_{y}b_{0}%
$. We can solve for eigen values
\begin{equation}
E_{1\pm}^{\prime}=\frac{\varepsilon_{d}+\varepsilon_{p}}{2}+\sqrt
{(\frac{\varepsilon_{d}-\varepsilon_{p}}{2})^{2}+4V^{\prime2}\pm V^{\prime
2}{\big |}%
{\displaystyle\sum\limits_{l}}
\cos2\delta\theta_{l}e^{2i\overrightarrow{k}\cdot\overrightarrow{r}_{l1}%
}{\big |}} \tag{A-3a}%
\end{equation}%
\begin{equation}
E_{2\pm}^{\prime}=\frac{\varepsilon_{d}+\varepsilon_{p}}{2}-\sqrt
{(\frac{\varepsilon_{d}-\varepsilon_{p}}{2})^{2}+4V^{\prime2}\pm V^{\prime
2}{\big |}%
{\displaystyle\sum\limits_{l}}
\cos2\delta\theta_{l}e^{2i\overrightarrow{k}\cdot\overrightarrow{r}_{l1}%
}{\big |}} \tag{A-3b}%
\end{equation}

\begin{equation}
E=\varepsilon_{p} \tag{A-3c}%
\end{equation}
where $\varepsilon_{p}$\ is two-fold degenerate. The eigen vectors are
\begin{equation}
\psi_{1+,\overrightarrow{k}\boldsymbol{,}\overrightarrow{q}}^{\prime
}(\overrightarrow{r})=%
{\displaystyle\sum_{j}}
e^{i\overrightarrow{k}\cdot\overrightarrow{R}_{j}}A_{1+}^{\prime}{\Big [}%
\psi_{d}(\overrightarrow{r}-\overrightarrow{R}_{j})+%
{\displaystyle\sum\limits_{l}}
\frac{V^{\prime}(-1)^{l}e^{i\overrightarrow{k}\cdot\overrightarrow{r}_{l}}%
}{E_{1+}^{\prime}-\varepsilon_{p}}\psi_{pl}(\overrightarrow{r}-\overrightarrow
{R}_{j,l}){\Big ](}%
\begin{array}
[c]{c}%
\cos\theta_{j}\\
e^{i\phi}\sin\theta_{j}%
\end{array}
{\Big )} \tag{A-4a}%
\end{equation}%
\begin{equation}
\psi_{1-,\overrightarrow{k}\boldsymbol{,}\overrightarrow{q}}^{\prime
}(\overrightarrow{r})=%
{\displaystyle\sum_{j}}
e^{i\overrightarrow{k}\cdot\overrightarrow{R}_{j}}A_{1-}^{\prime}{\Big [}%
\psi_{d}(\overrightarrow{r}-\overrightarrow{R}_{j})+%
{\displaystyle\sum\limits_{l}}
\frac{V^{\prime}(-1)^{l}e^{i\overrightarrow{k}\cdot\overrightarrow{r}_{l}}%
}{E_{1-}^{\prime}-\varepsilon_{p}}\psi_{pl}(\overrightarrow{r}-\overrightarrow
{R}_{j,l}){\Big ](}%
\begin{array}
[c]{c}%
\cos\theta_{j}\\
e^{i\phi}\sin\theta_{j}%
\end{array}
{\Big )} \tag{A-4b}%
\end{equation}%
\begin{equation}
\psi_{2+,\overrightarrow{k}\boldsymbol{,}\overrightarrow{q}}^{\prime
}(\overrightarrow{r})=%
{\displaystyle\sum_{j}}
e^{i\overrightarrow{k}\cdot\overrightarrow{R}_{j}}A_{2+}^{\prime}{\Big [}%
\psi_{d}(\overrightarrow{r}-\overrightarrow{R}_{j})+%
{\displaystyle\sum\limits_{l}}
\frac{V^{\prime}(-1)^{l}e^{i\overrightarrow{k}\cdot\overrightarrow{r}_{l}}%
}{E_{2+}^{\prime}-\varepsilon_{p}}\psi_{pl}(\overrightarrow{r}-\overrightarrow
{R}_{j,l}){\Big ](}%
\begin{array}
[c]{c}%
\cos\theta_{j}\\
e^{i\phi}\sin\theta_{j}%
\end{array}
{\Big )} \tag{A-4c}%
\end{equation}%
\begin{equation}
\psi_{2-,\overrightarrow{k}\boldsymbol{,}\overrightarrow{q}}^{\prime
}(\overrightarrow{r})=%
{\displaystyle\sum_{j}}
e^{i\overrightarrow{k}\cdot\overrightarrow{R}_{j}}A_{2-}^{\prime}{\Big [}%
\psi_{d}(\overrightarrow{r}-\overrightarrow{R}_{j})+%
{\displaystyle\sum\limits_{l}}
\frac{V^{\prime}(-1)^{l}e^{i\overrightarrow{k}\cdot\overrightarrow{r}_{l}}%
}{E_{2-}^{\prime}-\varepsilon_{p}}\psi_{pl}(\overrightarrow{r}-\overrightarrow
{R}_{j,l}){\Big ](}%
\begin{array}
[c]{c}%
\cos\theta_{j}\\
e^{i\phi}\sin\theta_{j}%
\end{array}
{\Big )} \tag{A-4d}%
\end{equation}
where $A_{1\pm}^{\prime}$\ and\ $A_{2\pm}^{\prime}$\ are normalization
constants.\ One can easily see the similarity between the wave functions with
or without bond-bending. The polarization can be computed as eq. (13). The
result is
\begin{equation}
\overrightarrow{P}_{\overrightarrow{q}}^{\prime}=\left\langle
-e\overrightarrow{r}\right\rangle \approx\frac{4\lambda V^{\prime}}%
{\pi(\varepsilon_{p}-\varepsilon_{d})\Delta E_{cf}^{\prime}}(eI\widehat{e}%
_{z})[\cos\phi\sin(a_{0}q_{x}/2)+\sin\phi\sin(b_{0}q_{y}/2)] \tag{A-5}%
\end{equation}
where $\Delta E_{cf}^{\prime}$ is the energy difference between t$_{2g}$
states and e$_{g}$ states and%
\begin{equation}
I=%
{\displaystyle\int}
d^{3}\overrightarrow{r}\,\psi_{yz}^{\ast}(\overrightarrow{r})z\psi
_{py}(\overrightarrow{r}-a_{0}\widehat{e}_{x}/2)=%
{\displaystyle\int}
d^{3}\overrightarrow{r}\,\psi_{zx}^{\ast}(\overrightarrow{r})z\psi
_{px}(\overrightarrow{r}-b_{0}\widehat{e}_{y}/2) \tag{A-6}%
\end{equation}

\subsubsection{References}

1. Z. J. Huang, Y. Cao, Y. Y. Sun, Y. Y. Xue and C. W. Chu, Phys. Rev.
\textbf{B56} 2623 (1997).

2. B. Lorentz, Y. Q. Wang, Y. Y. Sun and C. W. Chu, Phys. Rev. Lett.
\textbf{B70} 212412 (1997).

3. T. Kimura, T. Goto, H. Shintani, K. Ishizaka, T. Arima and Y. Tokura,
Nature \textbf{426}, 55 (2003).

4. T. Kimura, G. Lawes, T. Goto, Y. Tokura and A. P. Ramirez, Phys. Rev.
\textbf{B71}, 224425 (2005).

5. L. C. Chapon, G. R. Blake, M. J. Gutmann, S. Park, N. Hur, P.G. Radaelli,
and S-W. Cheong, Phys. Rev. Lett. \textbf{93}, 177402 (2004)

6. M. Kenzelmann, A. B. Harris, A. Aharony, O. Entin-Wohlman, T. Yildirim, Q.
Huang, S. Park, G. Lawes, C. Broholm, N. Rogado, R. J. Cava, K. H. Kim, G.
Jorge, and A. P. Ramirez, Phys. Rev. \textbf{B74}, 014429 (2006).

7. Y. Yamasaki, H. Sagayama, T. Goto1, M. Matsuura, K. Hirota, T. Arima and Y.
Tokura, arXiv:cond-matt/0701430v1 (2007).

8. M. Kenzelmann, G. Lawes, A.B. Harris, G. Gasparovic, C. Broholm, A.P.
Ramirez, G.A. Jorge, M. Jaime, S. Park, Q. Huang, A.Ya. Shapiro, and L.A.
Demianets, arXiv:cond-matt/0701426 (2007).

9. M. Kenzelmann, A. B. Harris, S. Jonas, C. Broholm, J. Schefer, S. B. Kim,
C. L. Zhang, S.-W. Cheong, O. P. Vajk, and J.W. Lynn, Phys. Rev. Lett.
\textbf{95}, 087206 (2005).

10. Maxim Mostovoy, Phys. Rev. Lett. \textbf{96,} 067601(2005).

11. I. E. Chupis, arXiv:cond-matt/0702636 (2007).

12. Ivan A. Sergienko and E. Dagotto, Phys. Rev. \textbf{73,} 094434 (2006).

13. Ivan A. Sergienko, Cengiz Sen, and Elbio Dagotto, Phys. Rev. Lett.
\textbf{97,} 227204 (2006).

14. E. R. S\'{a}nchez Guajardo, arXiv:cond-matt/0608300v1 (2006).

15. Seongsu Lee, A. Pirogov, Jung Hoon Han, J.-G. Park, A. Hoshikawa, and T.
Kamiyama, Phys. Rev. \textbf{71,} 180413 (2005).

16. Hosho Katsura, Naoto Nagaosa and Aleander V. Balatsky, Phys. Rev. Lett.
\textbf{95}, 057205 (2005).

17. Though the use of spin current may raise the issue of whether it is
well-defined in the presence of spin-orbit interaction, we nevertheless still
use this term\ as no ambiguity involved here.

18. I. Dzyaloshinskii, J. Phys. Chem. Solids \textbf{4}, 241 (1958).

19. T. Moriya, Phys. Rev. \textbf{120}, 91 (1960).

20. Y. Aharonov and A. Casher, Phys. Rev. Lett. \textbf{53,} 319 (1984).

21. Chenglong Jia, Shigeki Onoda, Naoto Nagaosa, and Jung Hoon Han, Phys. Rev.
\textbf{74,} 224444 (2006).

22. C. D. Hu, Phys. Rev. \textbf{75}, 172106 (2007).

23. Maxim Mostovoy, Phys. Rev. Lett. \textbf{94,} 137205(2005).

24. T. Kimura, S. Ishihara, H. Shintani, T. Arima, K. T. Takahashi, K.
Ishizaka, and Y. Tokura, Phys. Rev. \textbf{B68,} 060403 (2003).

25. J. C.Slater and G. F. Koster, Phys. Rev. \textbf{94}, 1498 (1954).

26. T. Kimura, S. Ishihara, H. Shintani, T. Arima, K. T. Takahashi, K.
Ishizaka and Y. Tokura, Phys. Rev. \textbf{B68}, 060403 (2003).

27. According to the calculation of ref. 21, introducing the on-site Coulomb
repulsion will not change qualitatively the physical, picture.

28. B. W. A. Leurs, Z. Nazario, D.I. Santiago, J. Zaanen, cond-mat.arXiv:str-el/0705.2953v1.

\newpage

Figure caption

\bigskip

Fig. 1 Schematic helical spin configuration. $\phi$ is the angle between the
projections of spins on xy-plane and x-axis.\ The angle between spins and
z-axis is twice of $\theta_{j,m}=\overrightarrow{q}\cdot\overrightarrow
{R}_{j,m}/2$ defined in text.

Fig. 2 (a) Bond angle $\alpha$. (b) A two-dimensional lattice with
bond-bending. The solid dots and circles denote the transition metal ions and
oxygen atoms respectively. The atoms are labeled so as to facilitate later deduction.

Fig. 3 Polarization $P_{z}$ versus $\phi$ and $\phi_{q}$ where $\phi_{q}$\ is
the angle between $\overrightarrow{q}$\ and x-axis.\ 

Fig. 4 A spin current along the Mn-O-Mn bond (y-component of spins moving in
x-direction.) It can be coupled to the z-component of electric field and
induce polarization $P_{z}$.\ 

\newpage\newpage

Fig. 1%

\begin{figure}
[ptb]
\begin{center}
\includegraphics[
height=4.5065in,
width=3.128in
]%
{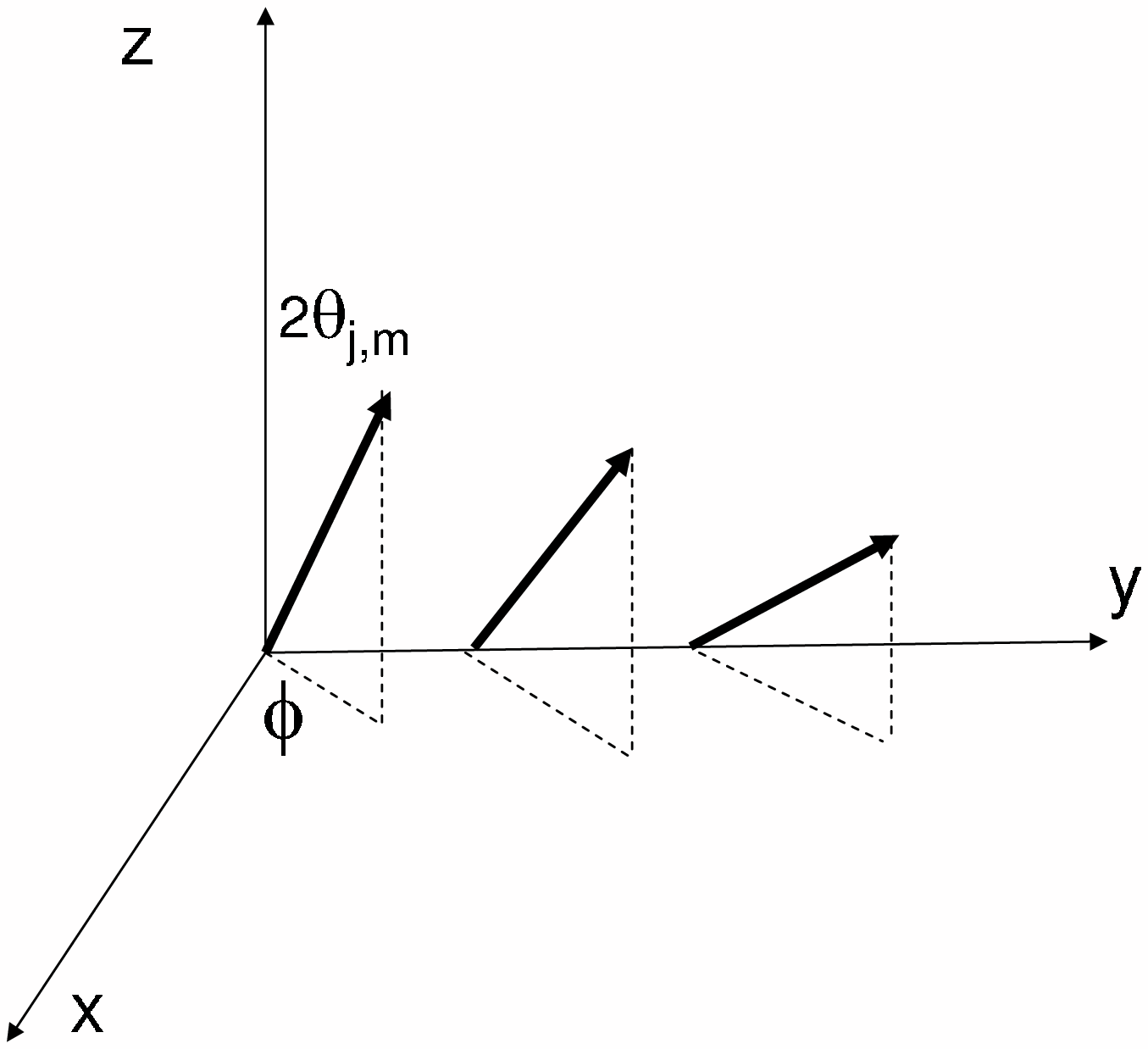}%
\end{center}
\end{figure}
\newpage

Fig. 2a\bigskip%

\begin{figure}
[ptb]
\begin{center}
\includegraphics[
height=3.6538in,
width=2.7449in
]%
{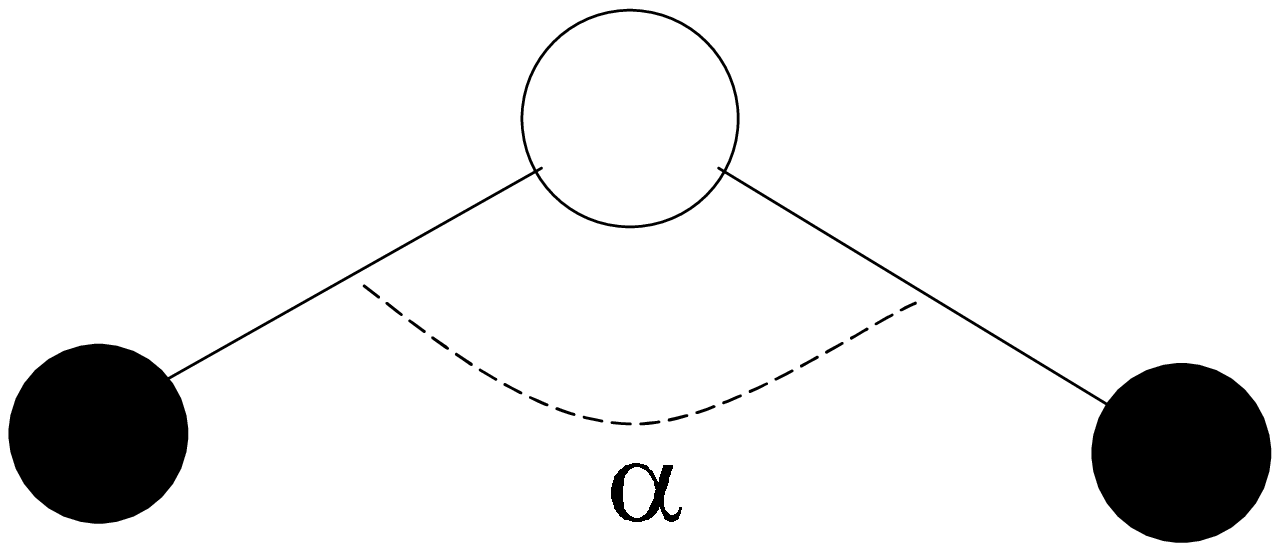}%
\end{center}
\end{figure}
\newpage

Fig. 2b%

\begin{figure}
[ptb]
\begin{center}
\includegraphics[
height=3.6538in,
width=2.7449in
]%
{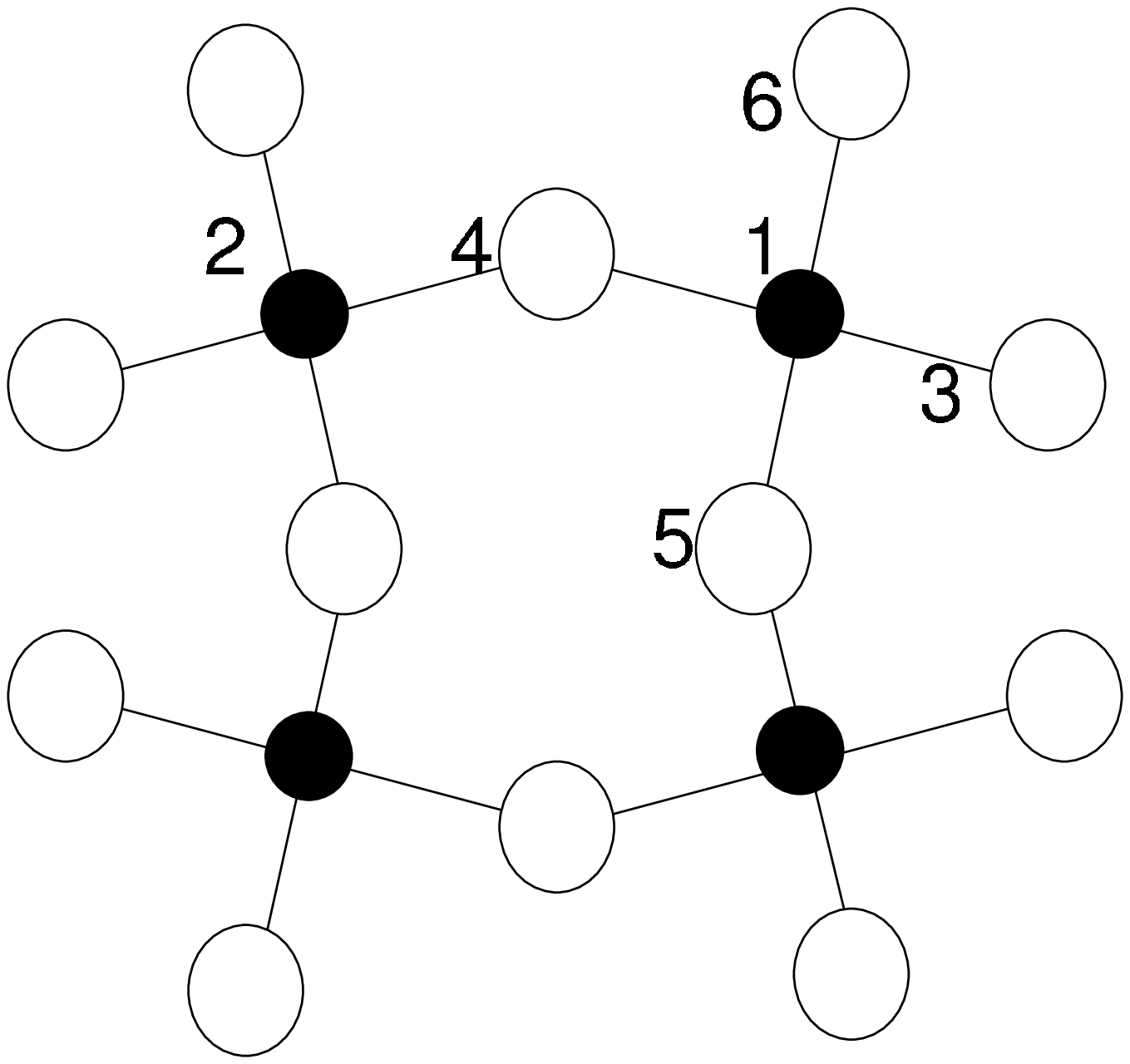}%
\end{center}
\end{figure}
\newpage

Fig. 3\bigskip%

\begin{figure}
[ptb]
\begin{center}
\includegraphics[
height=3.6538in,
width=4.171in
]%
{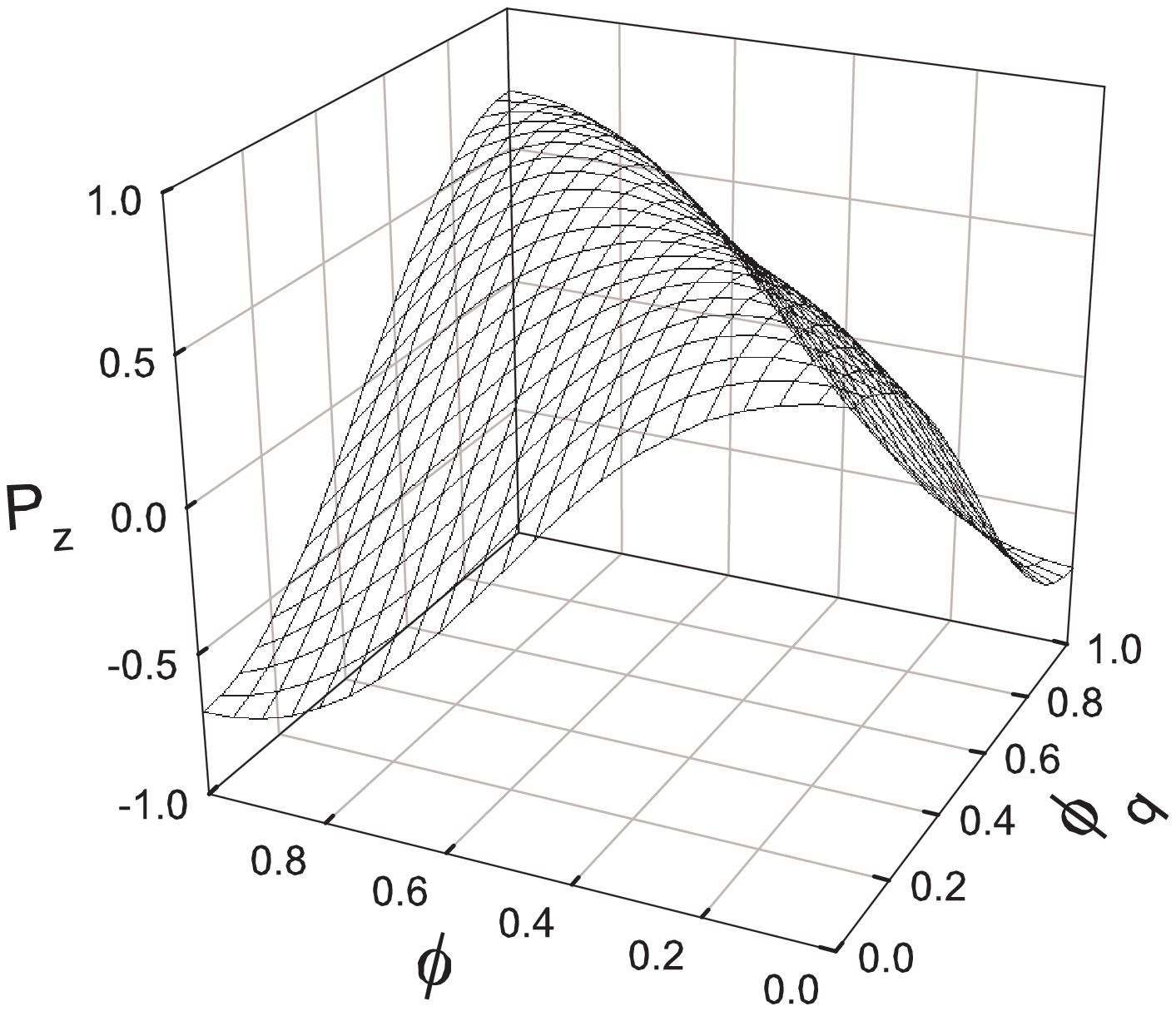}%
\end{center}
\end{figure}
\newpage

Fig. 4\bigskip%

\begin{figure}
[ptb]
\begin{center}
\includegraphics[
height=6.2353in,
width=4.8879in
]%
{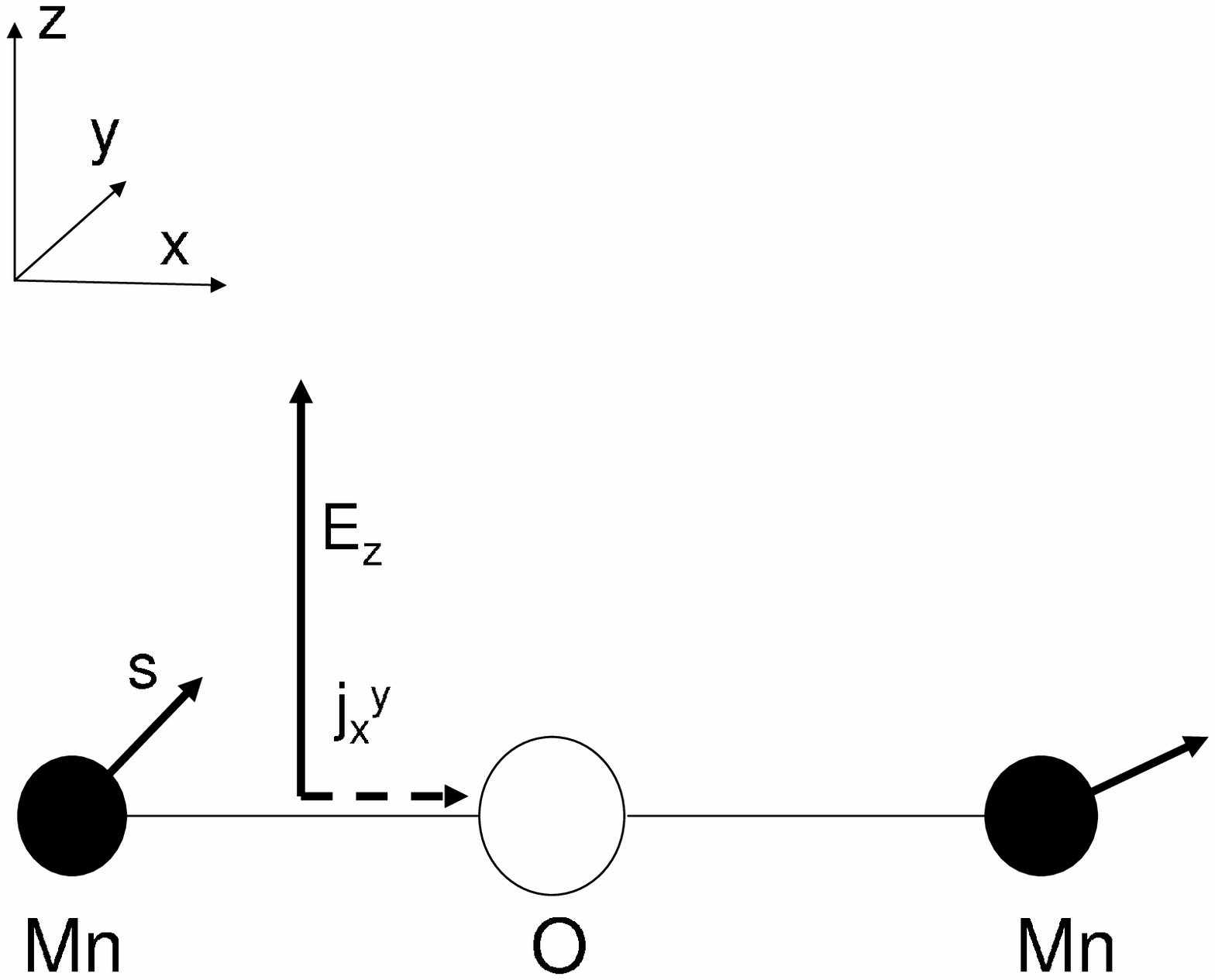}%
\end{center}
\end{figure}

\end{document}